\documentstyle[psfig]{laa-s}  
\newcommand{\reduceme}{\mbox{R\raisebox{-0.35ex}{E}D%
\hspace{-0.05em}\raisebox{0.85ex}{uc}\hspace{-0.90em}%
\raisebox{-.35ex}{{m}}\hspace{0.05em}E}}
\newcommand{\onen}{\makebox[0in][r]{1}}

\begin{document}
\title{Reliable random error estimation in the measurement of line-strength 
indices}
%
\author{N. Cardiel\inst{1}, J. Gorgas\inst{1}, J. Cenarro\inst{1} 
\and J.J. Gonz\'{a}lez\inst{2}}
%
%
\institute{Departamento de Astrof\'{\i}sica, Facultad de F\'{\i}sicas,
Universidad Complutense, 28040 Madrid, Spain
\and Instituto de Astronom\'{\i}a, U.N.A.M., Apdo. Postal 70-264, 04510
M\'{e}xico D.F., M\'{e}xico}
%
%
%
\abstract{
We present a new set of accurate formulae for the computation 
of random errors in the measurement of atomic and molecular line-strength 
indices. The new expressions are in
excellent agreement with numerical simulations. We have found that, in some
cases, the use of approximated equations can give misleading 
line-strength index errors. It is important to note that
accurate errors can only be achieved after a full control of the
error propagation throughout the data reduction with a parallel processing of
data and error images. Finally, simple recipes for the estimation of 
the required signal-to-noise ratio to achieve a fixed index error are
presented.
}
\keywords{methods: data analysis -- methods: analytical -- 
methods: statistical}
\maketitle
\markboth{Random errors in line-strength indices}{Cardiel et al.}
%
\section{Introduction}

Line-strength indices have proven to be excellent tools for the study of
absorption features in the spectra of astronomical objects. In particular,
since the pioneering work of Faber (1973), the behaviour of indices in
composite stellar systems has supplied fundamental clues for understanding
the history of their stellar populations (e.g. Burstein et al. 1984; Bica
et al. 1990; Gorgas et al. 1990; Worthey et al.  1992; Bender et
al. 1993; Jones \& Worthey 1995; Davies 1996).  At the light of up-to-date
stellar population models, line-strength indices can provide constrains about
important parameters such as mean age, metallicity, or abundance ratios (e.g.
Worthey 1994; Vazdekis et al. 1996; Bressan et al. 1996).  In addition,
line-strength indices have been widely employed to determine basic stellar
atmospheric parameters (Rich 1988; Zhou 1991; Terndrup et al. 1995) and star
cluster abundances (Brodie \& Hanes 1986; Mould et al. 1990; Gregg 1994;
Minniti 1995; Huchra et al. 1996), as well as to confront cosmological issues
by providing new distance indicators (Dressler et al. 1987) and a powerful
tool to investigate the redshift evolution of galaxies (Hamilton 1985;
Charlot \& Silk 1994; Bender et al. 1996).

Obviously, the suitability of line-strength indices to investigate the above
items relies on a proper determination of the associated index errors. For
illustration, it is interesting to note how errors in the measurement of
absorption features translate into uncertainties in the derived mean age and
metallicity of an old stellar population.  Taking the predictions of the
single-burst stellar population models of Worthey (1994) as a reference,
typical errors of 0.20 \AA\ in the Lick indices Fe5270 and H$\beta$ (see
below) translate into $\Delta{\rm [Fe/H]} \simeq 0.18$ dex and $\Delta{\rm
age} \simeq 6$ Gyr respectively (for a composite population of 10~Gyr and
metallicity around [Fe/H]=0.3).  Another example in which it is extremely
important to obtain reliable index errors is the analysis of the intrinsic
scatter of relations such as that of Mg$_2$ with velocity dispersion in
elliptical galaxies (Schweizer et al. 1990, Bender et al. 1993).

An accurate error determination is therefore needed in order to draw
confident interpretations from observed data. Although this requires the
estimation of both random and systematic errors, in this paper we will
concentrate in the former.  Whereas random errors can be readily derived by
applying statistical methods, systematic errors do not always allow such a
straight approach.  Several authors have dealt with the most common sources
of systematic errors in the measurement of line-strength indices. These
include: flux calibration effects (Gonz\'{a}lez 1993; Davies et al. 1993;
Cardiel et al. 1995), spectral resolution and velocity dispersion corrections
(Gorgas et al.  1990; Gonz\'{a}lez 1993; Carollo et al. 1993; Davies et al.
1993; Fisher et al. 1996; Vazdekis et al. 1997), sky subtraction
uncertainties (Saglia et al. 1993; Davies et al. 1993; Cardiel et al. 1995;
Fisher et al. 1995), scatter light effects (Gonz\'{a}lez 1993), wavelength
calibration and radial velocity errors (Cardiel et al. 1995; Vazdekis et al.
1997), seeing and focus corrections (Thomsen \& Baum 1987; Gonz\'{a}lez 1993;
Fisher et al. 1995), deviations from linearity response of the detectors
(Gorgas et al. 1990; Cardiel et al. 1995), and contamination by nebular
emission lines (Gonz\'{a}lez 1993; Goudfrooij \& Emsellem 1996).

Although it is always possible to obtain estimates of index errors by
performing multiple observations of the same object, analytical formulae to
avoid such an observing time effort are clearly needed.  In this paper we
present a set of analytical expressions to derive reliable errors in the
measurement of line-strength indices (hereafter we use {\it error\/} to quote
random errors exclusively).  General index definitions are given in
Section~2. Section~3 describes how a proper treatment of error propagation
throughout data reduction is a prerequisite to compute confident index
errors. Random errors can be obtained through numerical simulations,
presented in Section~4, or analytically. Previous works are briefly reviewed
in Section~5, with special emphasis to the approach followed by Gonz\'{a}lez
(1993). Our final set of formulae is presented in Section~6.  Section~7 gives
some recipes to estimate the required signal-to-noise ratios to achieve a
fixed index error.

\section{Definitions}

The strengths of absorption spectral features have been measured in different
ways so far.  However, although with slight differences among them, most
authors have employed line-strength indices with definitions close to the
classical expression for an equivalent width:
\begin{equation}
\label{eqerr:equivalentwidth}
W_{\lambda} ({\rm\AA}) = \int_{\rm line} \left( 1 - S(\lambda)/C(\lambda)
  \right) \; {\rm d}\lambda
\end{equation}
where $S(\lambda)$ is the observed spectrum and $C(\lambda)$ is the local
continuum usually obtained by interpolation of $S(\lambda)$ between two
adjacent spectral regions (e.g. Faber 1973; Faber et al.  1977; Whitford \&
Rich 1983; Gorgas 1987; Brodie \& Huchra 1990; Gonz\'{a}lez 1993; Rose
1994).  In order to avoid subjective determinations of the local continuum,
line-strength indices following Eq.~(\ref{eqerr:equivalentwidth}), referred
as {\it atomic indices\/}, are completely characterized by three wavelength
regions (bandpasses).  The spectral feature of interest is covered by the
central bandpass whereas the other two bandpasses, located towards the red
and blue of the central region, are employed to define the continuum
reference level through a linear interpolation. As pointed out by Geisler
(1984) (see also Rich 1988), at low spectral resolution a pseudo-continuum is
measured instead of a true continuum.

Line-strength indices are sometimes measured in magnitudes using:
\begin{equation}
  I({\rm mag}) = -2.5 \; \log_{10} \left( 
     1 - \frac{W_{\lambda}({\rm\AA})}{\Delta \lambda}
                             \right)
\end{equation}
where $\Delta \lambda$ is the width of the central bandpass.  These
line-strengths, referred as {\it molecular indices\/} since they are used for
molecular-band features, are defined with the help of broad bandpasses with
the continuum regions located far from the central feature.  On the other
hand, atomic indices, which measure the absorption of atomic spectral
features, have narrower and neighboring bandpasses.

Throughout this
paper we use the definitions given by Gonz\'{a}lez (1993), in which atomic
($I_{\rm a}$) and molecular ($I_{\rm m}$) indices are defined as follows: 
\begin{eqnarray}
\label{eqerr:iatom}
I_{\rm a}      & \equiv &   \int_{\lambda_{c_1}}^{\lambda_{c_2}}
  \left( 1- S(\lambda)/C(\lambda) \right) \; {\rm d}\lambda \\
\label{eqerr:imol}
I_{\rm m}      & \equiv &   -2.5 \; \log_{10} 
  \frac{\int_{\lambda_{c_1}}^{\lambda_{c_2}} 
  S(\lambda)/C(\lambda) \: {\rm d}\lambda}
  {\lambda_{c_2}-\lambda_{c_1}} 
\end{eqnarray}
where $\lambda_{c_1}$ and $\lambda_{c_2}$ are the limits of the central
bandpass (in \AA). The local pseudo-continuum $C(\lambda)$ 
is derived by
\begin{equation}
\label{eqerr:cont}
C(\lambda) \equiv S_b \frac{\lambda_r - \lambda}{\lambda_r - \lambda_b}
 + S_r \frac{\lambda - \lambda_b}{\lambda_r - \lambda_b} 
   \;\;\;\; {\rm where} \\
\end{equation}
\begin{equation}
\label{eqerr:fbluered}
S_b \equiv  
  \frac{\int_{\lambda_{b_1}}^{\lambda_{b_2}} S(\lambda) \: {\rm d}\lambda}
  {(\lambda_{b_2}-\lambda_{b_1})}, \qquad 
S_r \equiv  
  \frac{\int_{\lambda_{r_1}}^{\lambda_{r_2}} S(\lambda) \: {\rm d}\lambda}
  {(\lambda_{r_2}-\lambda_{r_1})} \\
\end{equation}
\begin{equation}
\label{eqerr:lambdabr}
\lambda_b  \equiv (\lambda_{b_1}+\lambda_{b_2})/2 , \qquad
\lambda_r  \equiv (\lambda_{r_1}+\lambda_{r_2})/2
\end{equation}
being $\lambda_{b_1}$, $\lambda_{b_2}$, $\lambda_{r_1}$, and $\lambda_{r_2}$
the limits of the blue and red bandpasses respectively. 

Although simplified versions of these expressions (i.e. considering a
constant continuum flux along the central bandpass, or replacing the
integrals by mean values) yield similar results at intermediate resolution,
the more accurate formulae must be favoured in order to guarantee the
comparisons with high-resolution high-S/N (signal-to-noise ratio) spectra
(specially for asymmetric indices, as already noted by Worthey et al. 1994).

Probably, the most widely index definition system employed so far is that
established by the Lick group (Burstein et al. 1984, 1986; Faber et al. 1985;
Gorgas et al. 1993; Worthey et al. 1994). In Table~1 we list the exact
definitions (as given by Trager 1997) for the 21 indices which constitute
the extended Lick system (see also Gonz\'{a}lez 1993). This table also
includes the definitions given by D\'{\i}az et al. (1989) for the Ca{\sc II}
triplet in the near-infrared. In this paper we will concentrate on the
analysis of line-strength errors for these particular indices, although the
derived analytical expressions are valid for any general index following
Eq.~(\ref{eqerr:iatom}) or~(\ref{eqerr:imol}).

Another interesting spectral feature which will be studied in this paper
is the amplitude of the $\lambda 4000$-\AA\ break
(D$_{4000}$). We adopt here the definition given by Bruzual (1983): 
\begin{equation}
\label{eqerr:d4000lambdanu}
{\rm D}_{4000} \equiv 
  \frac{\int_{4050}^{4250} S(\nu) \: {\rm d}\lambda}
       {\int_{3750}^{3950} S(\nu) \: {\rm d}\lambda}
\end{equation}
This {\it index\/} 
can be considered as a pseudo-color, being the combination of $S(\nu)$ and
$d\lambda$ due to historical reasons.

\begin{table*}
\caption[ ]{Bandpass definitions for the Lick
index system ---using the revised bandpass limits as given by Trager (1997)---,
together with the Ca{\sc II} triplet according to
D\'{\i}az et al. (1989).
The S/N constants $c_1$, $c_2$ and $c_3$ are explained in Section~7.}
\begin{flushleft}
\begin{tabular}{lcccc}
\hline
Index Name  & Central Bandpass (\AA) & Continuum Bandpasses (\AA) & 
\multicolumn{2}{c}{$c_i$} \\ \hline
            &                    &                    &         &        \\
\multicolumn{3}{c}{\raisebox{0mm}[0mm][3mm]{\sl Atomic Indices}}  & 
c$_1$   & c$_2$  \\
Ca4227   & 4222.250--4234.750 & 4211.000--4219.750 &        4.604  & 0.3684 \\
         &                    & 4241.000--4251.000 &               &        \\
G4300    & 4281.375--4316.375 & 4266.375--4282.625 &        8.537  & 0.2439 \\
         &                    & 4318.875--4335.125 &               &        \\
Fe4383   & 4369.125--4420.375 & 4359.125--4370.375 & {\onen}3.220  & 0.2580 \\
         &                    & 4442.875--4455.375 &               &        \\
Ca4455   & 4452.125--4474.625 & 4445.875--4454.625 &        7.038  & 0.3128 \\
         &                    & 4477.125--4492.125 &               &        \\
Fe4531   & 4514.250--4559.250 & 4504.250--4514.250 & {\onen}1.299  & 0.2511 \\
         &                    & 4560.500--4579.250 &               &        \\
Fe4668   & 4634.000--4720.250 & 4611.500--4630.250 & {\onen}7.757  & 0.2059 \\
         &                    & 4742.750--4756.500 &               &        \\
H$\beta$ & 4847.875--4876.625 & 4827.875--4847.875 &        7.301  & 0.2539 \\
         &                    & 4876.625--4891.625 &               &        \\
Fe5015   & 4977.750--5054.000 & 4946.500--4977.750 & {\onen}6.455  & 0.2158 \\
         &                    & 5054.000--5065.250 &               &        \\
Mg$_b$   & 5160.125--5192.625 & 5142.625--5161.375 &        8.032  & 0.2472 \\
         &                    & 5191.375--5206.375 &               &        \\
Fe5270   & 5245.650--5285.650 & 5233.150--5248.150 &        9.250  & 0.2313 \\
         &                    & 5285.650--5318.150 &               &        \\
Fe5335   & 5312.125--5352.125 & 5304.625--5315.875 & {\onen}0.741  & 0.2685 \\
         &                    & 5353.375--5363.375 &               &        \\
Fe5406   & 5387.500--5415.000 & 5376.250--5387.500 &        7.256  & 0.2893 \\
         &                    & 5415.000--5425.000 &               &        \\
Fe5709   & 5696.625--5720.375 & 5672.875--5696.625 &        6.362  & 0.2679 \\
         &                    & 5722.875--5736.625 &               &        \\
Fe5782   & 5776.625--5796.625 & 5765.375--5775.375 &        6.134  & 0.3067 \\
         &                    & 5797.875--5811.625 &               &        \\
NaD      & 5876.875--5909.375 & 5860.625--5875.625 &        8.113  & 0.2496 \\
         &                    & 5922.125--5948.125 &               &        \\
Ca1      & 8483.000--8513.000 & 8447.500--8462.500 &        8.852  & 0.2951 \\
         &                    & 8842.500--8857.500 &               &        \\
Ca2      & 8527.000--8557.000 & 8447.500--8462.500 &        8.330  & 0.2777 \\
         &                    & 8842.500--8857.500 &               &        \\
Ca3      & 8647.000--8677.000 & 8447.500--8462.500 &        7.750  & 0.2583 \\
         &                    & 8842.500--8857.500 &               &        \\
         &                    &                    &               &        \\
\multicolumn{3}{c}{\raisebox{0mm}[0mm][3mm]{\sl Molecular Indices}} & 
\multicolumn{2}{c}{c$_3$ } \\
CN$_1$   & 4142.125--4177.125 & 4080.125--4117.625 & 
\multicolumn{2}{c}{0.2241} \\
         &                    & 4244.125--4284.125 & 
\multicolumn{2}{c}{      } \\
CN$_2$   & 4142.125--4177.125 & 4083.875--4096.375 & 
\multicolumn{2}{c}{0.2691} \\
         &                    & 4244.125--4284.125 & 
\multicolumn{2}{c}{      } \\
Mg$_1$   & 5069.125--5134.125 & 4895.125--4957.625 & 
\multicolumn{2}{c}{0.1662} \\
         &                    & 5301.125--5366.125 & 
\multicolumn{2}{c}{      } \\
Mg$_2$   & 5154.125--5196.625 & 4895.125--4957.625 & 
\multicolumn{2}{c}{0.1933} \\
         &                    & 5301.125--5366.125 & 
\multicolumn{2}{c}{      } \\
TiO$_1$  & 5936.625--5994.125 & 5816.625--5849.125 & 
\multicolumn{2}{c}{0.1824} \\
         &                    & 6038.625--6103.625 & 
\multicolumn{2}{c}{      } \\
TiO$_2$  & 6189.625--6272.125 & 6066.625--6141.625 & 
\multicolumn{2}{c}{0.1568} \\
         &                    & 6372.625--6415.125 & 
\multicolumn{2}{c}{      } \\
\hline
\end{tabular}
\end{flushleft}
\end{table*}
\section{Propagation of errors throughout data reduction}

The aim of the reduction process is to minimize the influence of data
acquisition imperfections on the estimation of the desired astronomical
quantity (see Gilliland 1992 for a short review on noise sources and
reduction processes of CCD data).  For this purpose, one must perform
appropriate manipulations with the data and calibration frames.  The
arithmetic work involved in this process must be taken into account in order
to get reliable estimates of line-strength errors.

In order to trace in full detail the error propagation, error frames must be
created at the beginning of the reduction procedure. After this point, error
and data frames should be processed in parallel, translating the basic
arithmetic manipulations performed over the data images into the error frames
by following the law of combination of errors.

The starting point is the creation of initial error images with
the expected RMS variances from photon statistics and read-out noise. For a
single spectrum:
\begin{equation}
 \sigma^2 [j] = \frac{1}{g} N_{\rm c} [j] + \sigma_{\rm RN}^2 [j]
 \label{eqerr:sninitial}
\end{equation}
where $\sigma^2 [j]$ is the variance in the pixel $[j]$ ($\sigma [j]$ in
number of counts, ADU, ---analogic to digital number---), $g$ the gain of the
A/D converter (in $e^{-}$/ADU), $N_{\rm c}[j]$ the number of counts in the
pixel $[j]$ (after the bias-level subtraction), and $\sigma_{\rm RN} [j]$ is
the read-out noise (in ADU)\footnote{Note that the apparent dimensional
inconsistency of Eq.~\ref{eqerr:sninitial} is not real, and arises from the
fact that one of the properties of the Poisson distribution is that its
variance is numerically equal to the mean expected number of events.}.

Some of the reduction steps that may have a non negligible effect in the
index errors are flatfielding, geometrical distortion corrections, wavelength
calibration, sky subtraction and rebinning of the spectra. Note that if error
spectra were computed from the final number of counts in the reduced data
frame, index errors would tend to be underestimated.  The extra benefit of a
parallel processing of data and error frames is the possibility of obtaining,
at any time of the reduction, the variation in the mean S/N ratio produced by
a particular reduction step. Under these conditions, it is possible to
determine which parts of the reduction process are more sensitive to errors
and even to decide whether some manipulations of the data images can be
avoided, either because the resulting S/N ratio is seriously reduced and the
benefit of the product insignificant, or because such manipulations become a
waste of time.

A full error propagation in the reduction of spectroscopic data has been
previously implemented by Gonz\'{a}lez (1993), and it is also included in the
reduction package \reduceme\ (Cardiel \& Gorgas 1997)\footnote{Available at:\\
http://www.ucm.es/OTROS/Astrof/reduceme/reduceme.html}.

To obtain errors on line-strength indices from the reduced data and error
spectra, two different approaches can be followed. On one hand, analytical
formulae to evaluate index errors as a function of the data and variance
values in each pixel can be applied (Gonz\'{a}lez 1993).  Another method is
to simulate numerically the effect of the computed pixel variances in the
index measurements (e.g. Cardiel et al. 1995).  Both techniques are examined
in the next sections.

\section{Numerical simulations}

The effect of random noise in the spectra of astronomical objects can be
simulated by introducing in each pixel Gaussian noise computed as:
\begin{equation}
  {\cal R}[j] = \sqrt{2} \times \sigma[j] \sqrt{-\ln(1-r_1)} 
                \cos(2 \pi r_2)
\end{equation}
where $\sigma^2[j]$ is the variance in the pixel $[j]$, and $r_1$ and $r_2$
two random numbers in the range $r_1,r_2 \in [0,1)$. After the creation of
$N_{\rm simul}$ synthetic spectra, index errors can be derived as the
unbiased standard deviation of the $N_{\rm simul}$ measurements of each index
(typically we employed $N_{\rm simul} \sim 1000$).

\begin{figure}
\centerline{\psfig{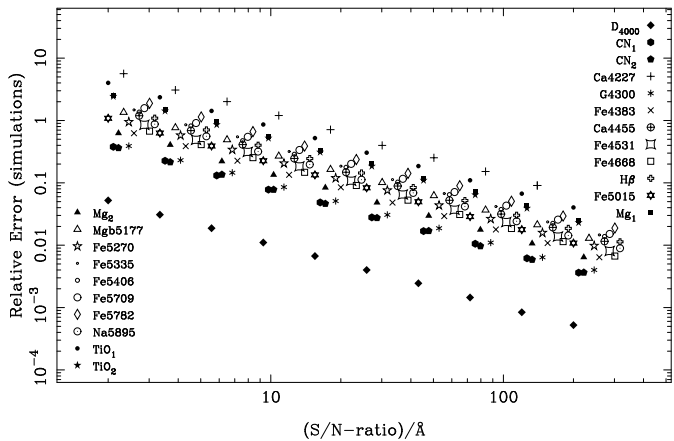}}
\caption[ ]{Relative errors from numerical simulations in the
measurement of 21 line-strength indices in the bright star HR~3428, as a
function of the mean S/N-ratio per \AA.}
\label{fig:simul_sn}
\end{figure}

For illustration, we show the results of numerical simulations using a high
S/N~ratio spectrum of the bright star HR~3428 as a template.  We have assumed
this spectrum to be noiseless, i.e. the nominal values of the measured
line-strength indices in this spectrum were considered as error free
reference values. By dividing this template spectrum by different constant
factors, we built a set of synthetic error spectra which were used to derive
index errors as a function of the mean (S/N-ratio)/\AA\ for different atomic
and molecular indices (Fig.~\ref{fig:simul_sn}).  In a logarithmic scale,
there is a clear linear correlation between the estimated relative error and
the (S/N-ratio)/\AA. In addition, and as it should be expected, at a fixed
(S/N-ratio)/\AA\ relative errors for atomic indices (with narrow bandpasses)
are larger than for molecular indices.

\section{Analytical formulae: previous works}

Several authors have employed analytical formulae to estimate errors in the
measurement of line-strength indices (Rich 1988; Brodie \& Huchra 1990;
Carollo et al. 1993; Gonz\'{a}lez 1993). However, the derived error
expressions have always been obtained from approximated versions of
Eqs.~(\ref{eqerr:iatom}) and~(\ref{eqerr:imol}), in which integrals are
replaced by averaged number of counts in the bandpasses. Although this
approximation works properly under certain circumstances, some additional
assumed simplifications seriously constrain the suitability of such formulae.
The expressions derived by Rich (1988) do not take into account readout noise
(which can be important at low count levels) nor the effect of the sky
subtraction.  In the formulae employed by Brodie \& Huchra (1990) and Carollo
et al.  (1993), although these factors are considered, important reduction
steps such as flatfielding (in particular slit illumination) are not taken
into account.  All these problems are readily settled with a parallel
reduction of data and error frames (Gonz\'{a}lez 1993, section~3 this
paper).  Although this approach requires a more elaborate reduction, it
yields the most confident results since it allows the application of
analytical formulae which consider the final error in each pixel of the
reduced spectra.

The most accurate set of analytic formulae published so far for the
computation of errors in line-strength indices are those presented by
Gonz\'{a}lez (1993). As a reference, we reproduce here his equations:
\begin{displaymath}
  \sigma[I_{\rm a}] = 
\end{displaymath}
\begin{equation}
  \label{eqerr:eiatom}
     \frac{S_c}{C_c} \sqrt{
     \left(\frac{\sigma_{S_c}}{S_c}\right)^2 +
     \frac{\sigma_{S_b}^2}{C_c^2} 
       \left( \frac{\lambda_r-\lambda_c}{\lambda_r-\lambda_b} \right)^2 +
     \frac{\sigma_{S_r}^2}{C_c^2} 
       \left( \frac{\lambda_b-\lambda_c}{\lambda_r-\lambda_b} \right)^2
     }
\end{equation}
\begin{equation}
  \label{eqerr:eimol}
  \sigma[I_{\rm m}] = \frac{2.5 \times 10^{0.4 I_{\rm m}}}
     {2.3026 (\lambda_{c_2}-\lambda_{c_1}) } \sigma[I_{\rm a}]
\end{equation}
where
\begin{eqnarray}
\label{eqerr:lambdac}
  \lambda_c & \equiv & (\lambda_{c_1}+\lambda_{c_2})/2 \\
  C_c       & \equiv & C(\lambda_c) \\
  S_c       & \equiv & \int_{\lambda_{c_1}}^{\lambda_{c_2}} 
    S(\lambda) \: d\lambda \\
  \label{eqerr:ejjgg1}
  (\sigma_{S_c}/S_c)^2 & \equiv & 1 /
    \int_{\lambda_{c_1}}^{\lambda_{c_2}} \frac{S^2(\lambda)}
    {\sigma^2(\lambda)} \: d\lambda \\
  \label{eqerr:ejjgg2}
  \sigma_{S_b}^2 & \equiv & S_b^2 /
    \int_{\lambda_{b_1}}^{\lambda_{b_2}} \frac{S^2(\lambda)}
    {\sigma^2(\lambda)} \: d\lambda \\
  \label{eqerr:ejjgg3}
  \sigma_{S_r}^2 & \equiv & S_r^2 /
    \int_{\lambda_{r_1}}^{\lambda_{r_2}} \frac{S^2(\lambda)}
    {\sigma^2(\lambda)} \: d\lambda
\end{eqnarray}
and
$\sigma^2(\lambda)$ is the variance of $S(\lambda)$ at the wavelength
$\lambda$.

\begin{figure}
\centerline{\psfig{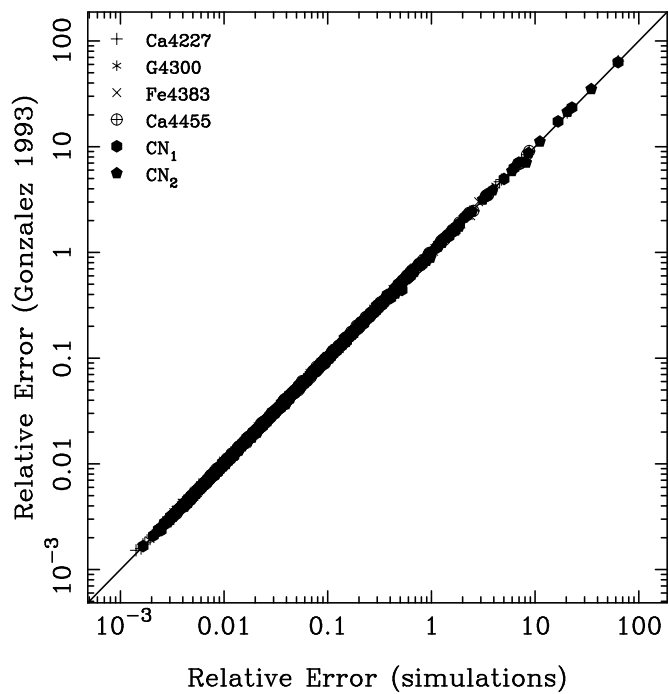}}
\caption[ ]{Comparison of the relative errors employing Gonz\'{a}lez's
formulae with those obtained from numerical simulations. The symbols
corresponds to the measurement of six different line-strength
indices in a sample of 350 stars from the Lick library.}
\label{fig:jjgg_simul}
\end{figure}

In order to check the accuracy of Eqs.~(\ref{eqerr:eiatom})
and~(\ref{eqerr:eimol}) we compared the predictions of such expressions with
the results obtained from numerical simulations.  We took spectra and
associated error spectra of a large homogeneous sample of 350 standard stars
from the Lick library (observed with RBS ---Richardson Brealy Spectrograph---
at the JKT ---Jacobus Kapteyn Telescope--- of the Roque de los Muchachos
Observatory, La Palma, February 1995).  The error spectra were obtained by
following a parallel reduction of data and error frames, as it has been
previously described. As it is apparent from Fig.~\ref{fig:jjgg_simul}, there
is an excellent agreement between Gonz\'{a}lez's formulae and the
simulations.

Unfortunately, the showed concordance between the analytical formulae and the
simulations can not be extrapolated to any general situation.  In particular,
one of the approximations used to derive Eqs.~(\ref{eqerr:eiatom})
and~(\ref{eqerr:eimol}) is the assumption that the mean signal-to-noise ratio
in each bandpass can be computed as the quadratic sum of the individual
signal-to-noise ratio in each pixel 
(Eqs.~\ref{eqerr:ejjgg1}--\ref{eqerr:ejjgg3}). In other words
\begin{equation}
\label{eqerr:snratio}
\frac{\left[ \sum_{i=1}^{N_{\rm pixels}} S(\lambda_{i}) \right]^2}
     {\sum_{i=1}^{N_{\rm pixels}} \sigma^2(\lambda_{i})} 
  \approx
\sum_{i=1}^{N_{\rm pixels}} \frac{S^2(\lambda_{i})}{\sigma^2(\lambda_{i})} 
\end{equation}
where $N_{\rm pixels}$ refers to the number of pixels involved in the
measurement of a particular bandpass.  The results presented in
Fig.~\ref{fig:jjgg_simul} were obtained employing data and error spectra in
which Eq.~(\ref{eqerr:snratio}) worked properly. 
However, it is straightforward to see that, in this
expression, a simultaneous combination of large and small $\sigma(\lambda)$
values would lead to a poor agreement between both terms.

\begin{figure}
\centerline{\psfig{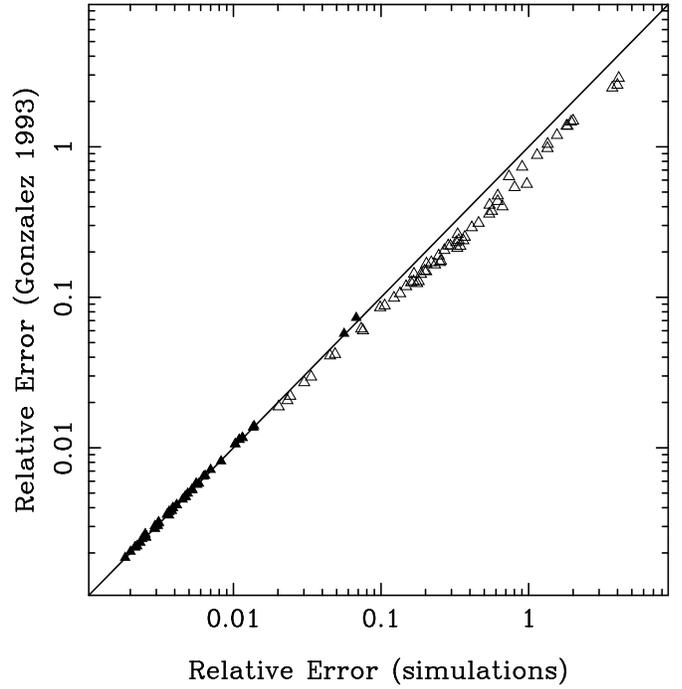}}
\caption[ ]{Comparison of the Mg$_2$ relative errors employing Gonz\'{a}lez's
formulae and numerical simulations. Filled triangles refer to a sample of 40
bright stars from the Lick Library, whereas open triangles correspond to
Mg$_2$ measurements along the radius of the cD galaxy in Abell~2255.  When
using Gonz\'{a}lez's expressions, there is a clear underestimation of the
relative errors in the galaxy due to the subtraction of a bright sky-line in
the Mg$_2$ central bandpass during the reduction process.}
\label{fig:mg2_abell2255}
\end{figure}

For example, discrepancies between Gonz\'{a}lez's formulae and numerical
simulations are apparent in the measurement of the Mg$_2$ index in the
spectra of the central dominant galaxy of the cluster Abell~2255 (observed
with the TWIN spectrograph at the 3.5$m$ Telescope of Calar Alto, August
1994), as it is shown in Fig.~\ref{fig:mg2_abell2255}. These differences are
due to the fact that a bright sky-line falls within the central bandpass of
the Mg$_2$ index. Sky subtraction during the reduction process introduces a
larger error in the pixels where sky-lines are present. As a result, the
reduced error spectra exhibit, simultaneously, pixels with very different
$\sigma(\lambda)$ values, and Eq.~(\ref{eqerr:snratio}) is a poor
approximation.

\section{The new formulae}

We have derived a more accurate set of
analytical formulae to compute errors in line-strength indices.
The expected random error
in the measurement of an atomic index is given by
\begin{equation}
  \sigma[I_{\rm a}] = \sigma 
  \left[ 
    \int_{\lambda_{c_1}}^{\lambda_{c_2}} \frac{S(\lambda)}{C(\lambda)} 
    {\rm d}\lambda
  \right]
\end{equation}
In practice, this integral must be transformed into the summation
\begin{equation}
  \sigma[I_{\rm a}] \simeq \sigma \left[ 
    \Theta \sum_{i=1}^{N_{\rm pixels}} 
    \frac{S(\lambda_i)}{C(\lambda_i)} \right] \equiv
    \Theta \; \sigma[{\cal I}]
\end{equation}
where $\Theta$ is the dispersion (in \AA/pixel), assuming a linear 
wavelength scale, and $N_{\rm pixels}$ the number of
pixels covering the central bandpass (note that, in general, fractions of
pixels must be considered in the borders of the bandpasses). ${\cal I}$ is a 
function of $2\times N_{\rm pixels}$ variables
${\cal I}(...,S(\lambda_i),...,C(\lambda_j),...)$ which verify
\begin{eqnarray}
  \label{eqerr:ccov1}
  {\rm cov}(S(\lambda_i),S(\lambda_j)) & = 0 & 
            \forall i,j \in [1,N_{\rm pixels}], \;\;\; i \neq j \\ 
  \label{eqerr:ccov2}
  {\rm cov}(S(\lambda_i),C(\lambda_j)) & = 0 & 
            \forall i,j \in [1,N_{\rm pixels}]                  \\ 
  \label{eqerr:ccov3}
  {\rm cov}(C(\lambda_i),C(\lambda_j)) & \neq 0 & 
            \forall i,j \in [1,N_{\rm pixels}], \;\;\; i \neq j  
\end{eqnarray}
since $C(\lambda)$ is computed from Eq.~(\ref{eqerr:cont}). Taking this
result into account:
\begin{displaymath}
  \sigma^2[{\cal I}] =
\end{displaymath}
\begin{displaymath}
    \sum_{i=1}^{N_{\rm pixels}} \left[
    \left( 
    \frac{\partial {\cal I}}{\partial S(\lambda_i)}
    \right)^2
    \sigma^2(\lambda_i) \right]
    +
    \sum_{i=1}^{N_{\rm pixels}} \left[ 
    \left( 
    \frac{\partial {\cal I}}{\partial C(\lambda_i)}
    \right)^2
    \sigma^2_{C({\lambda_i})} \right]
\end{displaymath}
\begin{displaymath}
    +
    \sum_{i=1}^{N_{\rm pixels}}
    \sum_{j=1, j \neq i}^{N_{\rm pixels}} \left[
    \left( 
    \frac{\partial {\cal I}}{\partial C(\lambda_i)}
    \right)
    \left( 
    \frac{\partial {\cal I}}{\partial C(\lambda_j)}
    \right)
    {\rm cov}(C({\lambda_i}),C({\lambda_j})) \right]
\end{displaymath}
\begin{displaymath}
    =
    \sum_{i=1}^{N_{\rm pixels}} \left[
    \frac{C^2(\lambda_i) \; \sigma^2(\lambda_i) + 
    S^2(\lambda_i) \; \sigma^2_{C({\lambda_i})}}
    {C^4({\lambda_i})}
    \right]
\end{displaymath}
\begin{equation}
    +
    \sum_{i=1}^{N_{\rm pixels}}
    \sum_{j=1, j \neq i}^{N_{\rm pixels}} \left[
    \left( 
    \frac{S(\lambda_i)}{C^2(\lambda_i)}
    \right)
    \left( 
    \frac{S(\lambda_j)}{C^2(\lambda_j)}
    \right)
    {\rm cov}(C({\lambda_i}),C({\lambda_j})) \right] 
\end{equation}
After some manipulation the covariance terms are
\begin{displaymath}
  {\rm cov}(C({\lambda_i}),C({\lambda_j})) = 
\end{displaymath}
\begin{equation}
    \langle C({\lambda_i}) C({\lambda_j}) \rangle -
    \langle C({\lambda_i}) \rangle 
    \langle C({\lambda_j}) \rangle =
\end{equation}
\begin{equation}
    \Lambda_1 \; \sigma^2_{S_b} +
    \Lambda_2 \; {\rm cov}(S_b,S_r) +
    \Lambda_3 \; {\rm cov}(S_r,S_b) +
    \Lambda_4 \; \sigma^2_{S_r}
\end{equation}
where we have defined the following four parameters
\begin{eqnarray}
  \Lambda_1 \equiv \frac{(\lambda_r-\lambda_i)(\lambda_r-\lambda_j)}
   {(\lambda_r-\lambda_b)^2} \\ 
  \Lambda_2 \equiv \frac{(\lambda_r-\lambda_i)(\lambda_j-\lambda_b)}
   {(\lambda_r-\lambda_b)^2} \\ 
  \Lambda_3 \equiv \frac{(\lambda_i-\lambda_b)(\lambda_r-\lambda_j)}
   {(\lambda_r-\lambda_b)^2} \\ 
  \Lambda_4 \equiv \frac{(\lambda_i-\lambda_b)(\lambda_j-\lambda_b)}
   {(\lambda_r-\lambda_b)^2}
\end{eqnarray}
Since $S_b$ y $S_r$ are not correlated, we obtain
\begin{equation}
  {\rm cov}(C({\lambda_i}),C({\lambda_j})) = 
    \Lambda_1 \; \sigma^2_{S_b} + \Lambda_4 \; \sigma^2_{S_r}
\end{equation}
and finally
\begin{displaymath}
\frac{\sigma^2[I_{\rm a}]}{\Theta^2} = 
  \sum_{i=1}^{N_{\rm pixels}} \left[
    \frac{C^2(\lambda_i) \; \sigma^2(\lambda_i) + 
    S^2(\lambda_i) \; \sigma^2_{C({\lambda_i})}}
    {C^4({\lambda_i})}
    \right] 
\end{displaymath}
\begin{equation}
\label{eqerr:eiatomfinal}
    +
    \sum_{i=1}^{N_{\rm pixels}}
    \sum_{j=1, j \neq i}^{N_{\rm pixels}} \left[
    \frac{S(\lambda_i) \; S(\lambda_j)}{C^2(\lambda_i) \; C^2(\lambda_j)}
    \left(
      \Lambda_1 \; \sigma^2_{S_b} + 
      \Lambda_4 \; \sigma^2_{S_r}
    \right)
    \right]
\end{equation}
where
\begin{equation}
  \sigma^2_{C(\lambda_i)} =
  \left( \frac{\lambda_r - \lambda_i}{\lambda_r - \lambda_b} \right)^2
  \sigma^2_{S_b} 
 + 
  \left( \frac{\lambda_i - \lambda_b}{\lambda_r - \lambda_b} \right)^2
  \sigma^2_{S_r} 
\end{equation}
\begin{equation}
  \sigma^2_{S_b} = 
  \frac{\Theta^2}{(\lambda_{b_2}-\lambda_{b_1})^2}
  \sum_{i=1}^{N_{\rm pixels(blue)}} \sigma^2(\lambda_i)
\end{equation}
\begin{equation}
  \sigma^2_{S_r} =
  \frac{\Theta^2}{(\lambda_{r_2}-\lambda_{r_1})^2}
  \sum_{i=1}^{N_{\rm pixels(red)}} \sigma^2(\lambda_i)
\end{equation}

Errors in the molecular indices are calculated through
\begin{equation}
\label{eqerr:eimolfinal}
\sigma[I_{\rm m}] = 2.5 \frac{\log_{10}{\rm e}}{10^{-0.4 I_{\rm m}}}
  \frac{1}{\lambda_{c_2}-\lambda_{c_1}} \sigma[I_{\rm a}]
\end{equation} 
which is identical to Eq.~(\ref{eqerr:eimol}).

\begin{figure}
\centerline{\psfig{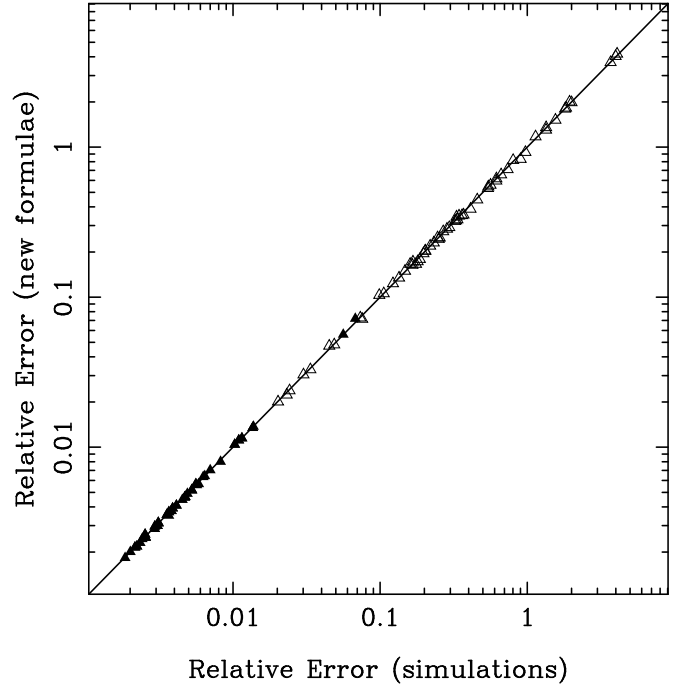}}
\caption[ ]{Comparison of the Mg$_2$ relative errors employing the new
formulae and numerical simulations for the samples shown in
Fig.~\ref{fig:mg2_abell2255}. The agreement between both methods is
complete.}
\label{fig:mg2_abell2255new}
\end{figure}

Using the new formulae, the computed index errors for the samples presented
in Figs.~\ref{fig:jjgg_simul} and~\ref{fig:mg2_abell2255} completely agree
with the results from numerical simulations (see 
Fig.~\ref{fig:mg2_abell2255new}).

The errors in the
4000~\AA\ break, defined in Eq.~(\ref{eqerr:d4000lambdanu}),
can be computed as
\begin{equation}
  \sigma^2[{\rm D}_{4000}] = 
    \frac{{\cal F}_r^2 \sigma_{{\cal F}_b}^2 + 
          {\cal F}_b^2 \sigma_{{\cal F}_r}^2}
    {{\cal F}_b^4}
\end{equation}
where
\begin{equation}
  {\cal F}_p \equiv
    \Theta
    \sum_{i=1}^{N_{\rm pixels}} \left[ \lambda_i^2 \;\; S(\lambda_i) \right]
\end{equation}
\begin{equation}
\sigma^2_{{\cal F}_p} = \Theta^2 \sum_{i=1}^{N_{\rm pixels}}
\left[ \lambda_i^4 \; \sigma^2(\lambda_i) \right]
\end{equation}
(the subscript $p$ refers indistinctly to $b$ or $r$).
Note that in this case ${\cal F}_b$ and ${\cal F}_r$ are not correlated.
D$_{4000}$ errors computed in this way show a perfect agreement with
numerical simulations.

\section{Estimation of S/N ratios}

As it is apparent from Fig.~\ref{fig:simul_sn}, there is a clear correlation
between the measured relative error and the (S/N-ratio)/\AA. Although such a
correlation arises naturally, quantitative estimates of errors as a function
of (S/N-ratio)/\AA\ are also expected to depend on index values.

We have shown that the use of approximate
formulae for the computation of errors can lead to misleading error
estimates. However, performing appropriate simplifications, it is possible to
derive simple expressions to estimate the absolute index error as a
function of the mean (S/N-ratio)/\AA. It can be shown that
\begin{equation}
\label{eqerr:iatom_estimation}
  \sigma[I_{\rm a}] \approx 
  \frac{c_1 - c_2 I_{\rm a}}{\rm SN(\AA)}
\end{equation}
where, to simplify, we have defined SN(\AA) as
\begin{equation}
  {\rm SN(\AA)} = \frac{1}{N \sqrt{\Theta}} \sum_{i=1}^{N} 
   \frac{S(\lambda_i)}{\sigma(\lambda_i)}
\end{equation}
(the summation extends over the three bandpasses, i.e. $N$ pixels).
The two constants $c_1$ and $c_2$ are defined as follows
\begin{equation}
    c_1 \equiv \Delta\lambda_c \; c_2
\end{equation}
\begin{equation}
    c_2 \equiv
    \sqrt{
      \frac{1}{\Delta\lambda_c}+
      \left(
      \frac{\lambda_r-\lambda_c}{\lambda_r-\lambda_b}
      \right)^2
      \frac{1}{\Delta\lambda_b}+
      \left(
      \frac{\lambda_c-\lambda_ b}{\lambda_r-\lambda_b}
      \right)^2
      \frac{1}{\Delta\lambda_r}
    }
\end{equation}
being $\Delta\lambda_b$, $\Delta\lambda_c$ and $\Delta\lambda_r$ the 
bandpass widths.

Similarly
\begin{equation}
\label{eqerr:imol_estimation}
  \sigma[I_{\rm m}] \approx 
  \frac{c_3}{\rm SN(\AA)}
\end{equation}
where
\begin{equation}
  c_3 \equiv 2.5 \; c_2 \; \log_{10} {\rm e}
\end{equation}

\begin{figure}
\centerline{\psfig{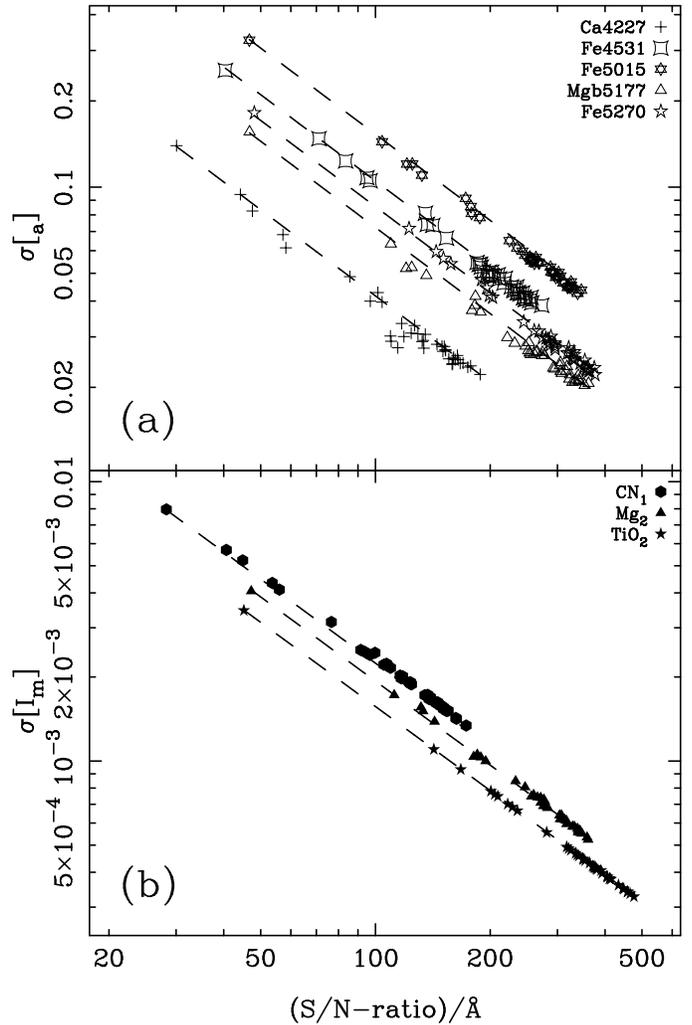}}
\caption[ ]{Absolute atomic (panel~a) and molecular (panel~b) line-strength
errors measured in the sample of 40 stars from
the Lick library as a function of the mean (S/N-ratio)/\AA. 
The predictions of
Eqs.~(\ref{eqerr:iatom_estimation}) and~(\ref{eqerr:imol_estimation}) are
plotted as dashed lines.}
\label{fig:errorestimation}
\end{figure}

Equations~(\ref{eqerr:iatom_estimation}) and~(\ref{eqerr:imol_estimation})
can be easily employed to predict the required (S/N-ratio)/\AA\ to achieve a
fixed index error. In Fig.~\ref{fig:errorestimation} we show the predictions
of these expressions for some particular indices compared with the actual
error measurements (from Eqs.~\ref{eqerr:eiatomfinal}
and~\ref{eqerr:eimolfinal}) in a star sample.  It is interesting to note that
the absolute error of molecular indices does not depend, in a first
approximation, on the absolute index value, although the contrary is true for
the atomic indices. This is the reason why a larger scatter is apparent in
panel~(a), where $I_{\rm a}$ in Eq.~(\ref{eqerr:iatom_estimation}) has been
replaced by its corresponding arithmetic mean in the sample.  Numerical values
for $c_1$, $c_2$ and $c_3$, which obviously depend on the considered index,
are given in Table~1.

Following the same procedure with the D$_{4000}$ index:
\begin{equation}
  \sigma[{\rm D}_{4000}] \approx
    \frac{{\rm D}_{4000}}{\sqrt{200}}
    \sqrt{
      \frac{1}{{\rm SN(\AA)}^2_b} +
      \frac{1}{{\rm SN(\AA)}^2_r}
    }
\end{equation}
where SN(\AA)$_b$ and SN(\AA)$_r$, 
the mean (S/N-ratio)/\AA\ in the blue and red band
respectively, will attain, in general, different values (given the large
wavelength coverage of the break). In this case, the relative D$_{4000}$
error does not depend on the absolute D$_{4000}$ value.

\section{Summary}

We present a set of analytical formulae to compute random errors in the
measurement of line-strength indices. The new expressions constitute an
improvement compared with the approaches followed in previous works,
providing a straight method to attain reliable random error estimates. In
particular, the derived equations should be preferred to methods involving
either numerical simulations (a computer time demanding method) or multiple
observations (usually prohibitive at low light levels). As a useful tool for
observation planning, we also provide simple recipes to estimate the required
signal-to-noise ratio to achieve a desired index error.

We want to stress here that full benefit from these formulae can only be
obtained after a proper treatment of the error propagation throughout the
data reduction. Apart from this, the analytical method followed to derived
the expressions given in Section~6 is exact, in the sense that it does not
involve any approximation (exception made for the unavoidable data
sampling).

\section{Acknowledgements}
We are grateful to the referee for useful comments. The JKT is operated on
the island of La Palma by the Royal Greenwich Observatory at the Observatorio
del Roque de los Muchachos of the Instituto de Astrof\'{\i}sica de Canarias.
The Calar Alto Observatory is operated jointly by the Max-Planck-Institute
f\"{u}r Astronomie, Heidelberg, and the Spanish Comisi\'{o}n Nacional de
Astronom\'{\i}a.  This work was supported by the Spanish ``Programa Sectorial
de Promoci\'{o}n General del Conocimiento'' under grant No. PB93-0456.


\end{document}